\pdfoutput=1

\documentclass[11pt,twoside,a4paper,cmspaper,final,collab]{cms-tdr}
\def\svnVersion{291187}\def\cmsCernNo{2013-037}\def\cmsCernDate{\today}\def\cmsMessage{Published in Physics Letters B as \href{http://dx.doi.org/10.1016/j.physletb.2015.04.045}{\doi{10.1016/j.physletb.2015.04.045.}}}

\begin{document}\cmsNoteHeader{EXO-12-052}

\hyphenation{had-ron-i-za-tion}
\hyphenation{cal-or-i-me-ter}
\hyphenation{de-vices}
\RCS$Revision: 286798 $
\RCS$HeadURL: svn+ssh://svn.cern.ch/reps/tdr2/papers/EXO-12-052/trunk/EXO-12-052.tex $
\RCS$Id: EXO-12-052.tex 286798 2015-04-29 13:42:13Z duggan $
\newlength\cmsFigWidth
\ifthenelse{\boolean{cms@external}}{\setlength\cmsFigWidth{0.98\columnwidth}}{\setlength\cmsFigWidth{0.6\textwidth}}
\newlength\cmsFigWidthTwo
\ifthenelse{\boolean{cms@external}}{\setlength\cmsFigWidthTwo{0.98\columnwidth}}{\setlength\cmsFigWidthTwo{0.49\textwidth}}
\ifthenelse{\boolean{cms@external}}{\providecommand{\cmsLeft}{top}}{\providecommand{\cmsLeft}{left}}
\ifthenelse{\boolean{cms@external}}{\providecommand{\cmsRight}{bottom}}{\providecommand{\cmsRight}{right}}
\cmsNoteHeader{EXO-12-052}
\title{Search for pair-produced resonances decaying to jet pairs in proton-proton collisions at $\sqrt{s}=8$\TeV}

\date{\today}
\abstract{
Results are reported of a general search for pair production of heavy resonances decaying
to pairs of hadronic jets in events with at least four jets.
The study is based on up to 19.4\fbinv of integrated luminosity from proton-proton
collisions at a center-of-mass energy of 8\TeV, recorded with the CMS detector at the LHC.
Limits are determined on the production of scalar top quarks (top squarks) in the framework of
R-parity violating supersymmetry and on the production of color-octet vector bosons (colorons).
First limits at the LHC are placed on top squark
production for two scenarios. The first assumes decay to a bottom quark and a light-flavor quark
and is excluded for masses between 200 and 385\GeV,
and the second assumes decay to a pair of light-flavor quarks and is excluded
for masses between 200 and 350\GeV at 95\% confidence level.
Previous limits on colorons decaying to light-flavor quarks are extended to exclude masses from
200 to 835\GeV.
}

\hypersetup{%
pdfauthor={CMS Collaboration},%
pdftitle={search for pair-produced resonances decaying to dijets at sqrt(s)= 8 TeV},%
pdfsubject={CMS},%
pdfkeywords={CMS, physics, dijets}}

\maketitle
\section{Introduction}
We present the results of a search for pair production of heavy resonances
decaying to pairs of light- and heavy-flavor quarks in
multijet events. The analysis is based on data samples corresponding to as
much as $19.4\pm0.5$\fbinv~\cite{lumi} of integrated luminosity
from proton-proton collisions at $\sqrt{s}=8$\TeV, collected with
the CMS detector~\cite{cmsdet} at the CERN LHC in 2012. Events that have at
least four jets with high transverse momentum (\pt) with respect to the beam
direction are selected and investigated for evidence of pair-produced dijet
resonances.

Many models of particle physics beyond the standard model (SM) incorporate particles that decay
into fully hadronic final states. Supersymmetric (SUSY) models are SM extensions,
which simultaneously solve the hierarchy problem and unify particle interactions~\cite{nilles,haber}.
In natural SUSY models, where there is minimal fine-tuning, the top quark
superpartner (top squark) and the superpartners of the Higgs boson (higgsinos) are
required to be light~\cite{natsusy,natsusy1,natsusy2,natsusy3,natsusy4}.
Natural SUSY is underconstrained
in certain R-parity violating (RPV) scenarios~\cite{rpvsusy}.
R-parity is a quantum number defined as $R=(-1)^{3B+L+2S}$, where
$B$ and $L$ are the baryon and lepton numbers, respectively, and $S$
is the spin.
The RPV superpotential, $W$, is defined as
\begin{equation}
W = \frac{1}{2} \lambda_{ijk} L_i L_j E_k^c + \lambda^{\prime}_{ijk} L_i Q_jD_k^c
+ \frac{1}{2} \lambda^{\prime \prime}_{ijk} U_i^c D_j^c D_k^c,
\end{equation}
where $\lambda$ are the couplings, $i$,$j$,$k$ are the generation
indices, $c$ is the charge conjugation, $L$ and $Q$ are the doublet superfields of the lepton and
quark, respectively, and $E$, $D$, and $U$ are the singlet superfields of
the lepton, down-type and up-type quarks, respectively. Models that
incorporate RPV may allow baryon number violation through a
non-zero $\lambda^{\prime \prime}_{\mathrm{UDD}}$ coupling, and one such
unconstrained scenario~\cite{hadronicstop} is that of the hadronically decaying top squark,
$\PSQt \to \PQq \PQq^{\prime}$.
If the top squarks are pair-produced in hadronic collisions and then decay via such an RPV
process, the final state would consist of four jets with no momentum imbalance in the
transverse plane.

In addition to top squark production, hadron collider searches for pair production of
resonances decaying into jet pairs are sensitive to a number of
models that predict new particles carrying color quantum numbers.
Some models predict pair production through gg interactions of
color-octet vectors, also called colorons (C)~\cite{colorons},
which then decay to quark pairs. The associated final state of
the signal is characterized by the presence of four high-\pt jets.

CDF collaboration has placed 95\% confidence level (CL) exclusion limits~\cite{cdfstops}
on top squark production followed by RPV decays in the mass range 50--90\GeV
and on coloron production in the mass range 50--125\GeV.
At the LHC, ATLAS has placed limits on scalar gluon
masses between 100 and 185\GeV~\cite{atlaspairs0}, and separately
for masses between 150 and 287\GeV~\cite{atlaspairs1}. The CMS search
for paired dijet resonances resulted in limits on coloron masses between
250 and 740\GeV~\cite{exo11016}. However, none of these searches has
been sensitive enough to set limits on hadronic RPV decays of directly
produced top squarks.

In this paper, we concentrate on searches for top squarks and colorons.
The benchmark signals are those where the top squark is the
lightest supersymmetric particle, and in one scenario decays into
two light quarks, and in the second scenario it decays into a b quark and a light
quark~\cite{tcgut,exoquarks,unichiral,altchiral,multijets,bstop}. We
separately consider the possibility of decays within the coloron model
($\Pg\Pg\to\mathrm{CC} \to \PQq\PAQq\PQq\PAQq$).

The analysis employs a well-established search strategy with optimized
event selections. The distribution of a variable representative of the top
squark mass is investigated for evidence of a signal consistent with
localized deviations from the estimated large, steeply falling SM background
to data. The estimate of the background is performed with a fit to the falling
part of the mass spectrum in data, and a SM MC analysis is used to
optimize the signal selection and to derive systematic uncertainties.

\section{CMS experiment}
The central feature of the CMS apparatus~\cite{cmsdet}
is a superconducting solenoid of 6\unit{m} internal diameter, providing a
magnetic field of 3.8\unit{T}. Within the superconducting solenoid volume
are a silicon pixel and strip tracker, a lead tungstate electromagnetic
calorimeter (ECAL), and a hadron calorimeter (HCAL), which is made of
interleaved layers of scintillator and brass absorber. Muons are measured
in gas ionization detectors embedded in the steel return yoke outside the
solenoid. Extended forward calorimetry complements the coverage provided
by the barrel and endcap detectors. Energy deposits from hadronic jets
are measured using the ECAL and HCAL.
A more detailed description of the CMS detector, together
with a definition of the coordinate system used and the relevant kinematic
variables, can be found in Ref.~\cite{cmsdet}.

\section{Triggering and object reconstruction}
One data set, representing 19.4\fbinv, was recorded over the entire 2012
data taking period with a multilevel trigger system,
which selected events with at least four jets with $\pt>80$\GeV to be reconstructed
from only calorimeter information. In addition, a second data set was recorded using the
same trigger logic, but with a lower jet \pt threshold. This threshold was decreased
progressively from 50 to 45\GeV during the 2012 data taking period.
The latter data represent only a subset of the entire 2012 data set,
corresponding to an integrated luminosity of 12.4\fbinv. The analysis is
separated into two parts: a dedicated ``low-mass'' search with a
focus on the mass region from 200 to 300\GeV, which takes advantage of this lower
jet \pt threshold, and a ``high-mass'' search focusing on top squark masses above
300\GeV, which uses the entire 19.4\fbinv data set and extends the expected
top squark mass search sensitivity by 40\GeV.

The analysis is based upon objects reconstructed using the CMS Particle Flow algorithm~\cite{pflow}.
This method combines calorimeter information with reconstructed charged particle tracks to
identify individual particles such as photons, leptons, and neutral
and charged hadrons. The energy of photons is directly
obtained from the calibrated ECAL measurement. The energy of the electron
is determined from a combination of its track momentum at the main
interaction vertex, the corresponding ECAL cluster energy, and the
energy sum of all bremsstrahlung photons associated to the track. The
energy of a muon is obtained from its associated track momentum. The
charged hadron energy is calculated from a combination of the track
momentum and the corresponding ECAL and HCAL energies, corrected for
zero-suppression effects, and calibrated for the combined response function
of the calorimeters. Finally, the energy of neutral hadrons is
obtained from the corresponding corrected ECAL and HCAL energies.
Jets are reconstructed from the particle flow ``objects'' using the
anti-\kt algorithm~\cite{ak5} with a distance parameter of 0.5
in $y$-$\phi$ space, where $y$ is the rapidity.

Jet energy scale corrections~\cite{jec} are applied to account for
the combined response function of the calorimeters to hadrons.
The corrections are derived from Monte
Carlo (MC) simulation and are confirmed with in situ measurements
of the energy balance of dijet and photon+jet events.
In data, a small
residual correction factor is included to account for
differences in jet response between data and simulation.
The total size of the applied corrections is approximately 5--10\%,
and the corresponding uncertainties vary from 3 to 5\%,
depending on the measured jet pseudorapidity $\eta$ and \pt.
To remove misidentified jets, which arise primarily from
calorimeter noise, jet quality criteria~\cite{jetid} are
applied. More than 99.8\% of all selected jets, in both
data and signal event samples, satisfy these criteria.

To identify jets produced by b quark hadronization, the
analysis uses the medium selection of the combined secondary
vertex b-tagging algorithm~\cite{btagger}. The algorithm
employs a multivariate technique, which takes as input
information from the transverse impact parameter with respect
to the primary vertex of the associated tracks and from
characteristics of the reconstructed secondary vertices.
The output of the algorithm is used to discriminate b quark
jets from light-flavor and gluon jets, with typical values
of b-tagging efficiency and misidentification probabilities
of 72\% and 1.1\%, respectively.

\section{Generation of simulated events}
Both top squark production and coloron production are simulated
using the
\MADGRAPH 5.1.5.12 \cite{madgraph}
event generator with the CTEQ6L1 parton distribution
functions~\cite{cteqPDFs}, and
their decays are simulated using the \PYTHIA~6.426~\cite{pythia} MC
program.
Top squark signal events are generated
with up to two additional initial-state partons, and each top squark
decays into two jets through
the $\lambda^{''}_{\mathrm{UDD}}$ quark RPV coupling. Two scenarios
are considered for this coupling. First, the coupling $\lambda^{''}_{312}$,
where the three numerical subscripts refer to the quark
generations of the corresponding quarks, is set to a non-zero
value such that the decay of the top squark to two light-flavor
jets is allowed. The second case instead sets a non-zero value for
$\lambda^{''}_{323}$, resulting in top squark decay into one b jet and
one light-flavor jet.
In both of the above cases, the branching fraction of the top squark decay to two jets
is set to 100\%.
For the generation of this signal, all superpartners except the top squarks
are taken to be decoupled~\cite{tcgut,exoquarks,unichiral,altchiral,multijets}
and no intermediate particles are produced in the top squark decay.
Top squarks are generated with masses from 100\GeV to 1\TeV
in 50\GeV steps for both coupling scenarios. The cross section
estimates~\cite{susyxsecuncert} are
made at next-to-leading order (NLO) with next-to-leading-logarithm (NLL)
corrections~\cite{susyxsec1, susyxsec2, susyxsec3, susyxsec4, susyxsec5},
and assigned appropriate theoretical uncertainties~\cite{susyxsecuncert}.
For the coloron signal scenario, we consider the case where each coloron
decays into two light-flavor jets with a branching fraction of 100\%. For
this signal, masses are generated from 100\GeV to 2\TeV, and NLO cross
section estimates are used. For both the top squark and coloron models,
the natural width of the signal resonance is taken to be much smaller
than the resolution of the detector. Backgrounds from SM multijet processes
are simulated through matched tree-level matrix elements for two- to
four-jet production using \MADGRAPH, and these events are showered
through \PYTHIA. In all samples, the MLM matching
procedure~\cite{matching} is used,
and simulation of the
CMS detector is performed with \GEANTfour~\cite{geant4}.

\section{Event selection}
\label{sec:evtsel}
Events recorded with the four-jet triggers are required
to have a well-reconstructed primary event vertex~\cite{goodPV}.
Events must also contain at least four jets, each with
$\abs{\eta} < 2.5$ and reconstructed \pt greater
than 80\GeV for the low-\pt trigger and
120\GeV for the higher-\pt trigger.
With the above requirements, the offline efficiency is above
99\% for all selected events.

\begin{figure}[hbt]
\centering
\includegraphics[width=\cmsFigWidthTwo]{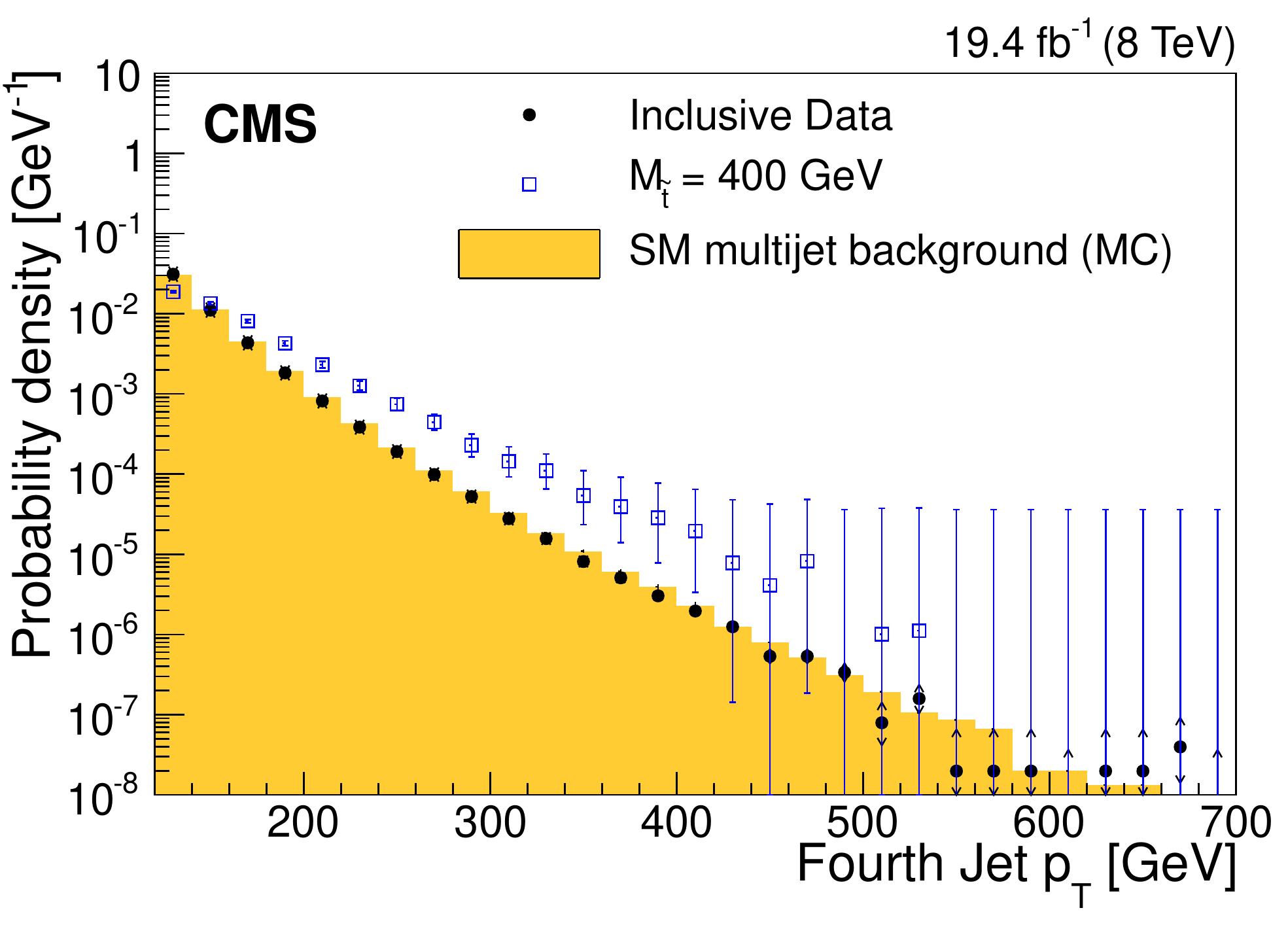}
\includegraphics[width=\cmsFigWidthTwo]{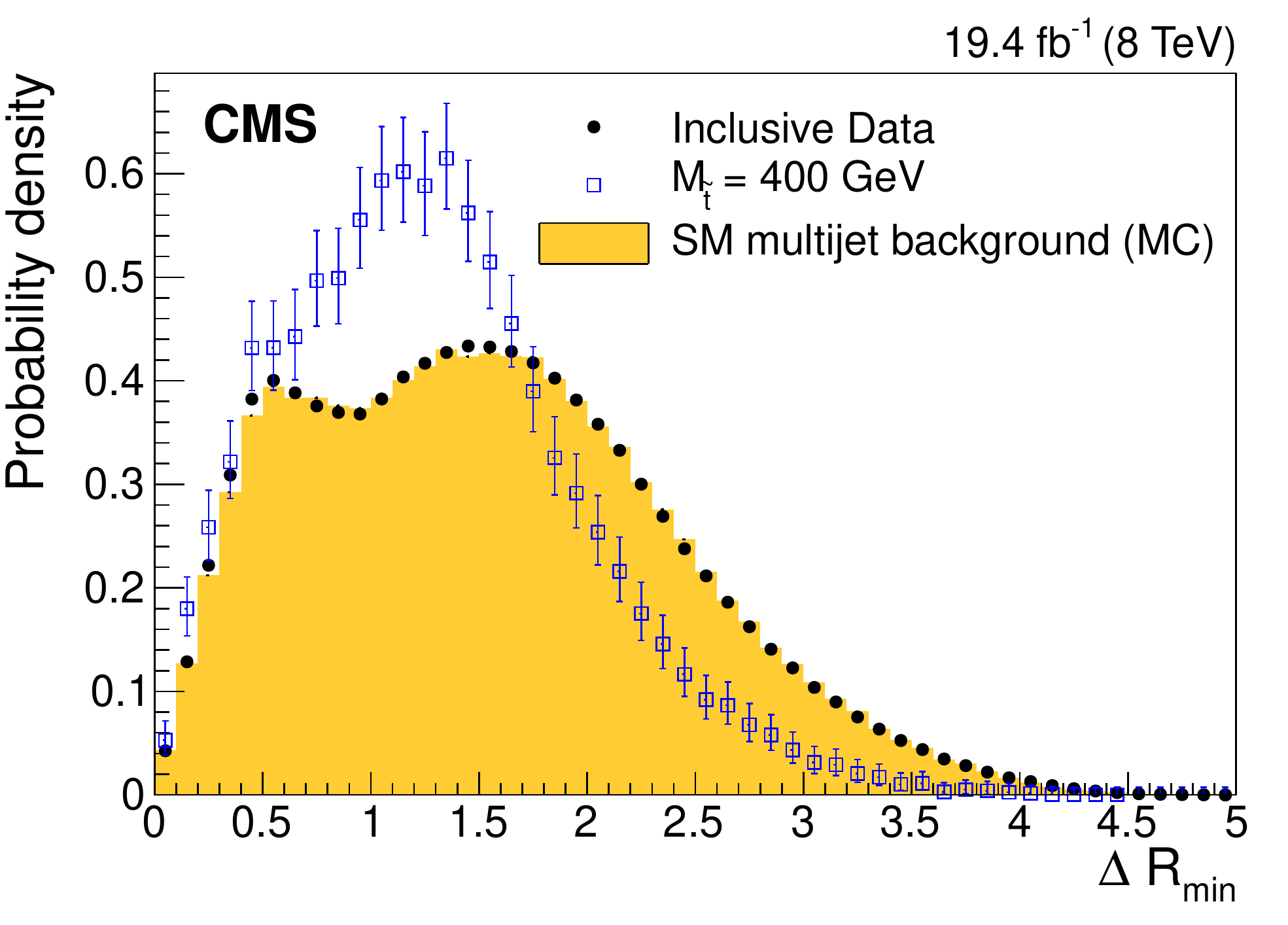}
        \caption{Probability density distributions of the fourth highest jet \pt (\cmsLeft)
                 and $\Delta R_{\text{min}}$ (\cmsRight) for
                 events from data, the simulated SM multijet
                 sample, and a 400\GeV top squark signal.
                 Statistical uncertainties are shown for the
                 top squark signal as vertical bars and for
                 data as arrows.
                 Events contain at least four jets, each with
                 $\pt>120$\GeV and $\abs{\eta} < 2.5$, and
                 all distributions have an area
                 normalized to unity.}
\label{fig:4thjetpt_dRmin}
\end{figure}

The leading four jets, ordered in \pt, are used to create three unique combinations
of dijet pairs per event. A distance variable is implemented to select
the jet pairing that best corresponds to the two resonance decays,
$\Delta R = \sqrt{\smash[b]{(\Delta\eta)^2 + (\Delta\phi)^2}}$, where
$\Delta\eta$ and $\Delta\phi$ are the differences in $\eta$ and $\phi$ of
between two the jets, respectively. This variable~\cite{dRpairing} exploits
the smaller relative distance between daughter jets from the same top squark parent
decays compared to that between uncorrelated jets. For each dijet
pair configuration the value of $\Delta R_{\text{dijet}}$ is calculated:
\begin{equation}
\Delta R_{\text{dijet}} = \sum_{i=1,2}\abs{\Delta R^{i} - 1},
\label{eqn:dRpairing}
\end{equation}
where $\Delta R^{i}$ represents the separation between two jets in dijet pair $i$.
An offset of 1 has been chosen since this maintains a maximal signal efficiency
while minimizing the selection of dijet systems composed of resolved jets from
radiated gluons paired with their parent jet. The configuration that minimizes the value
$\Delta R_{\text{dijet}}$ is selected, with $\Delta R_{\text{min}}$ representing the
minimum $\Delta R_{\text{dijet}}$ for the event. Figure~\ref{fig:4thjetpt_dRmin} shows
the probability density distributions of the fourth highest jet \pt and the $\Delta R_{\text{min}}$ variable
for data events, those of a simulated SM multijet sample, and those of 400\GeV top squark
signal sample.

\begin{figure}[hbt]
\centering
\includegraphics[width=\cmsFigWidthTwo]{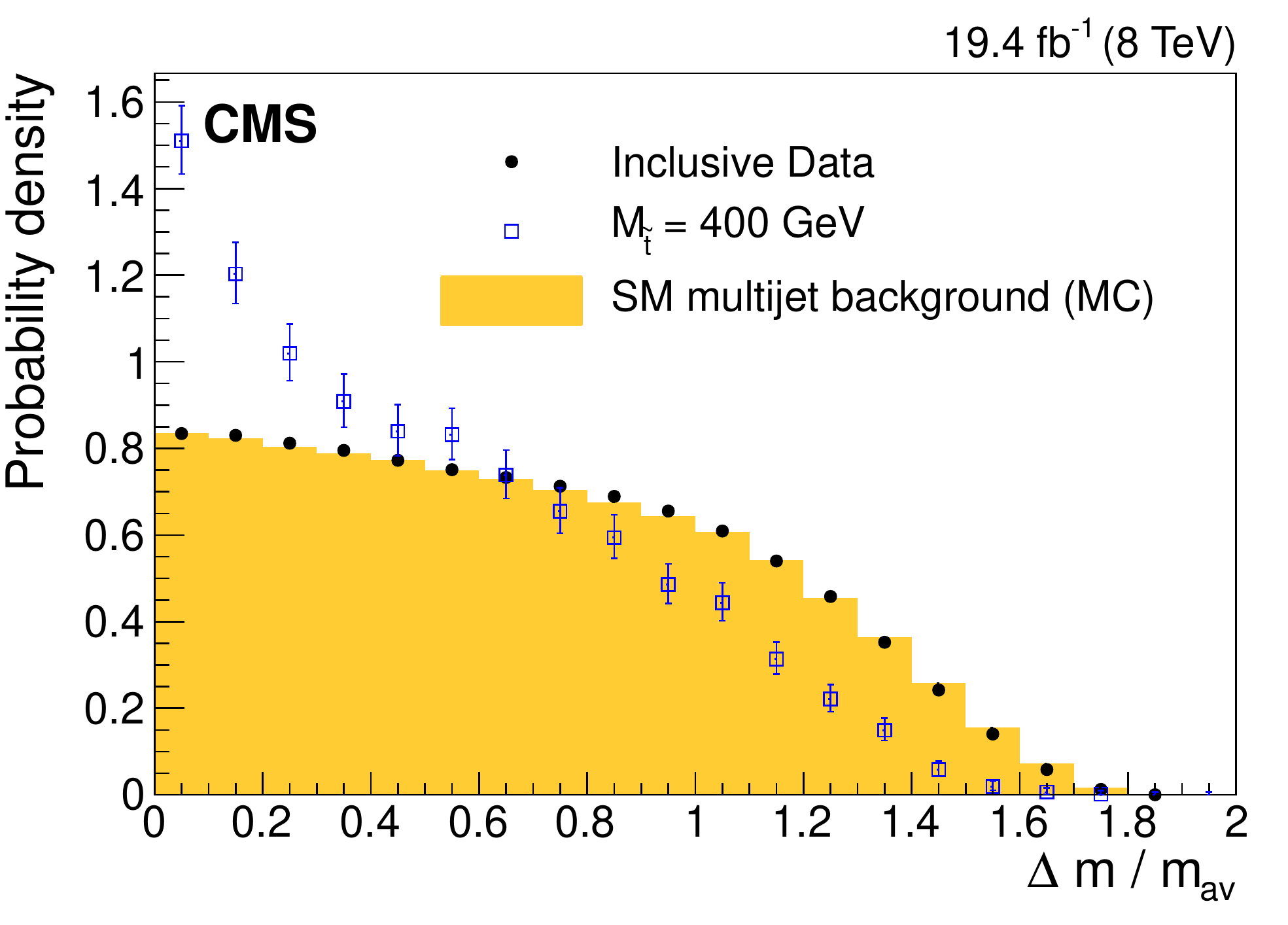}
\includegraphics[width=\cmsFigWidthTwo]{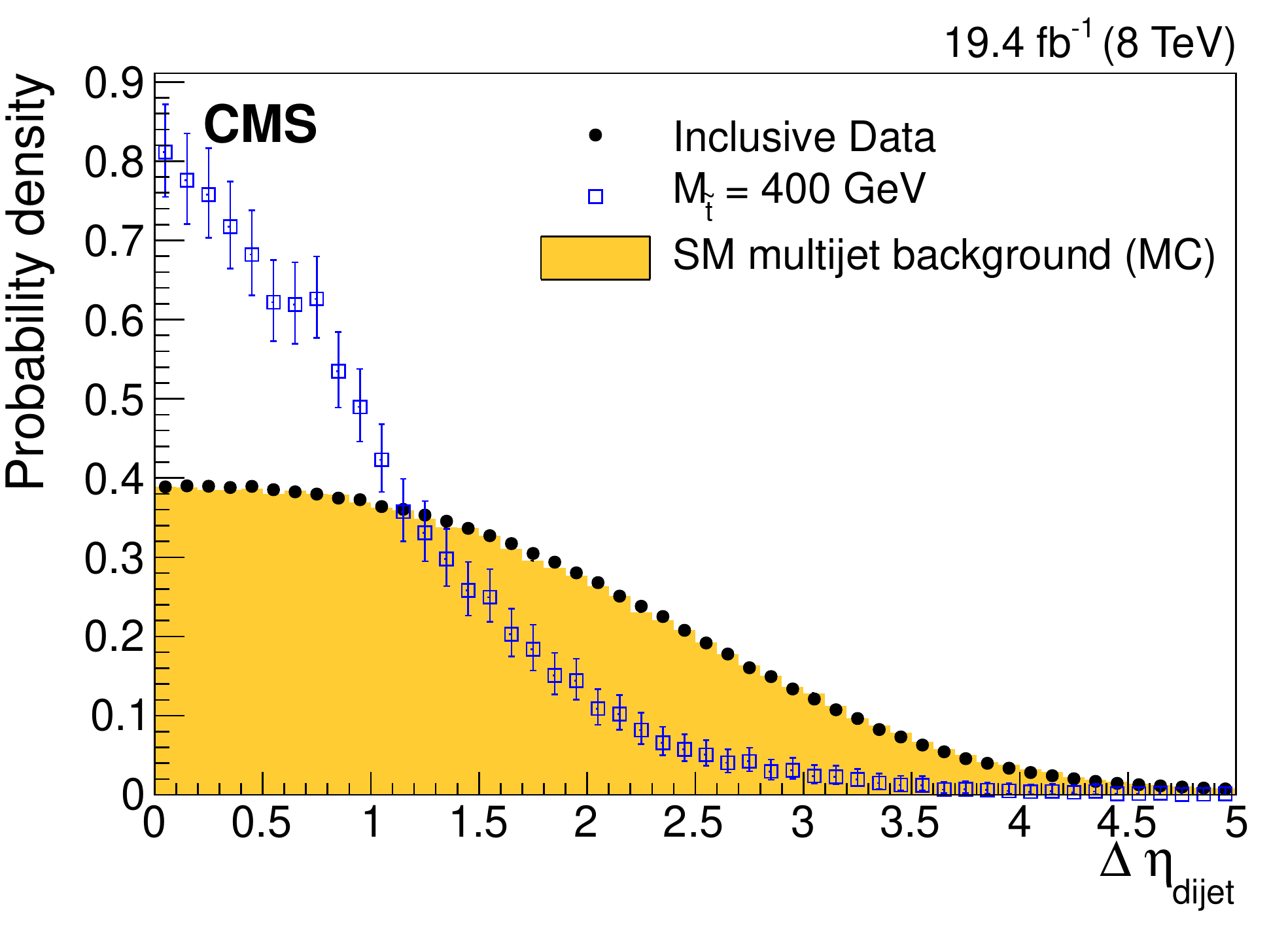}
        \caption{Probability density distributions of $\Delta m/m_{\text{av}}$ (\cmsLeft)
                 and $\Delta\eta_{\text{dijet}}$ (\cmsRight) for
                 events from data, the simulated SM multijet
                 sample, and a 400\GeV top squark signal.
                 Statistical uncertainties are shown for the
                 top squark signal as vertical bars and for
                 data as arrows.
                 Events contain at least four jets, each with
                 \pt$>120$\GeV and $\abs{\eta} < 2.5$, and
                 all distributions have an area normalized to unity.}
\label{fig:delta_m_eta}
\end{figure}

Once a dijet pair configuration is chosen, two additional quantities
are used to reject the backgrounds from SM multijet
events and incorrect signal pairings:
the pseudorapidity difference between the two dijet systems $\Delta\eta_{\text{dijet}}$,
and
the absolute value of the fractional mass difference $\Delta m/m_{\text{av}}$,
where $\Delta m$ is the difference between the two dijet masses and $m_{\text{av}}$
is their average value. In signal events where the correct pairing is chosen,
the $\Delta m/m_{\text{av}}$ quantity is peaked at zero with a much narrower distribution
than that for SM multijet background or incorrectly paired signal events.
Thus, the sensitivity of the search benefits from imposing a maximum value on
$\Delta m/m_{\text{av}}$. Similarly, it is
advantageous to require that $\Delta\eta_{\text{dijet}}$ be small.
Figure~\ref{fig:delta_m_eta} shows the probability density distributions of the $\Delta m/m_{\text{av}}$ and $\Delta\eta_{\text{dijet}}$
variables for data events, those of a simulated SM multijet sample,
and those of 400\GeV top squark signal sample.
An additional kinematic variable $\Delta$ is calculated for each dijet system:
\begin{eqnarray}
\Delta = \left(\sum_{i=1,2}\abs{\pt^{i}}\right) - m_{\text{av}},
\label{eqn:delta}
\end{eqnarray}
where the \pt sum is over the two jets in the dijet configuration.
This type of variable has been used extensively in hadronic resonance
searches at both the Tevatron and the LHC~\cite{cdfmultijets,cmsmultijets,gluino2011,exo11016,gluino2014}.
Requiring a minimum value of $\Delta$ results in a lowering of the
peak position value of the $m_{\text{av}}$ distribution from background SM multijet events.
With this selection the modeling of the background shape can be extended to lower values of $m_{\text{av}}$, making
a wider range of top squark and coloron masses accessible to the search.

Finally, as the presence of heavy-flavor final state jets is a natural extension of the
RPV top squark scenarios, the use of b tagging is exploited to further increase signal
sensitivity by increasing background rejection. We consider two scenarios: the
heavy-flavor search, which uses b tagging to increase the sensitivity for top squark decays
into heavy-flavor jets, and the inclusive search, which focuses instead on decays into
light-flavor jets.

\begin{figure}[hbt]
\centering
\includegraphics[width=\cmsFigWidthTwo]{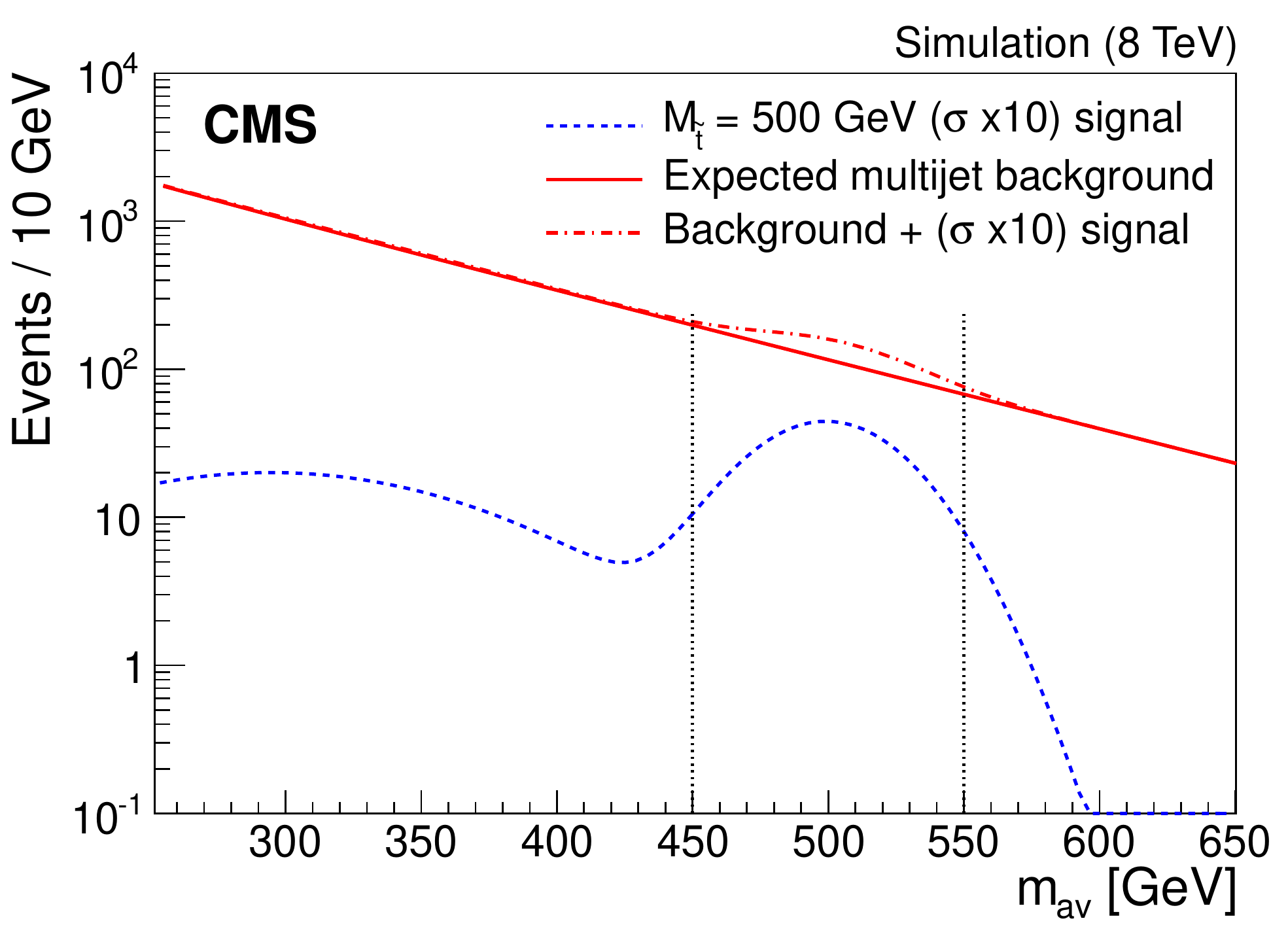}
        \caption{
        Distributions of the fit to simulated background SM multijet events (solid red line) and a 500\GeV top
        squark (dashed blue line), normalized to a factor of ten times its cross section,
        are shown for the high-mass optimization scenario. The dotted vertical lines represent
        the integration window used by the optimization procedure.}
\label{fig:sigbkgopt}
\end{figure}

\begin{figure}[hbtp]
\centering
\includegraphics[width=\cmsFigWidthTwo]{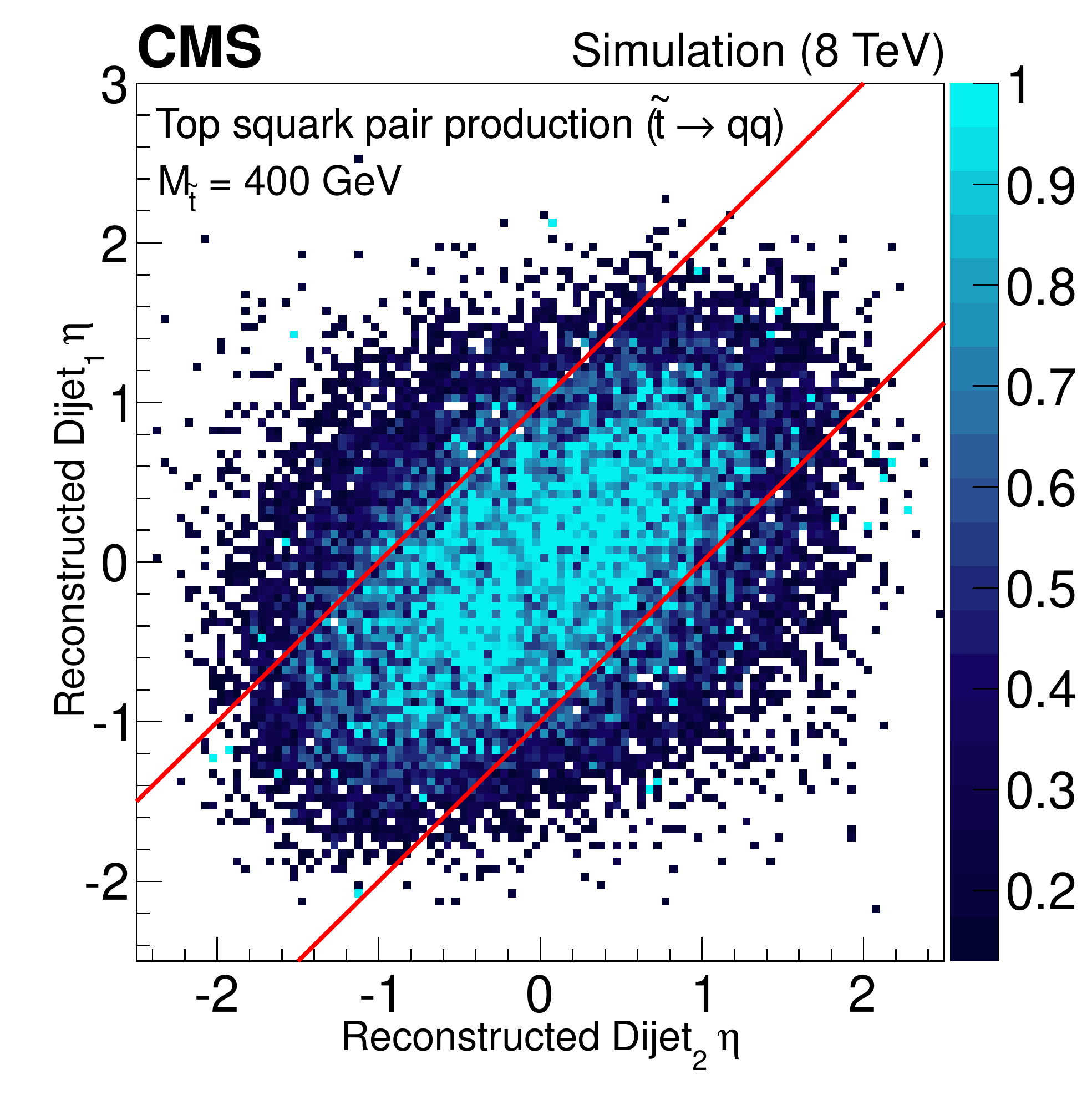}
\includegraphics[width=\cmsFigWidthTwo]{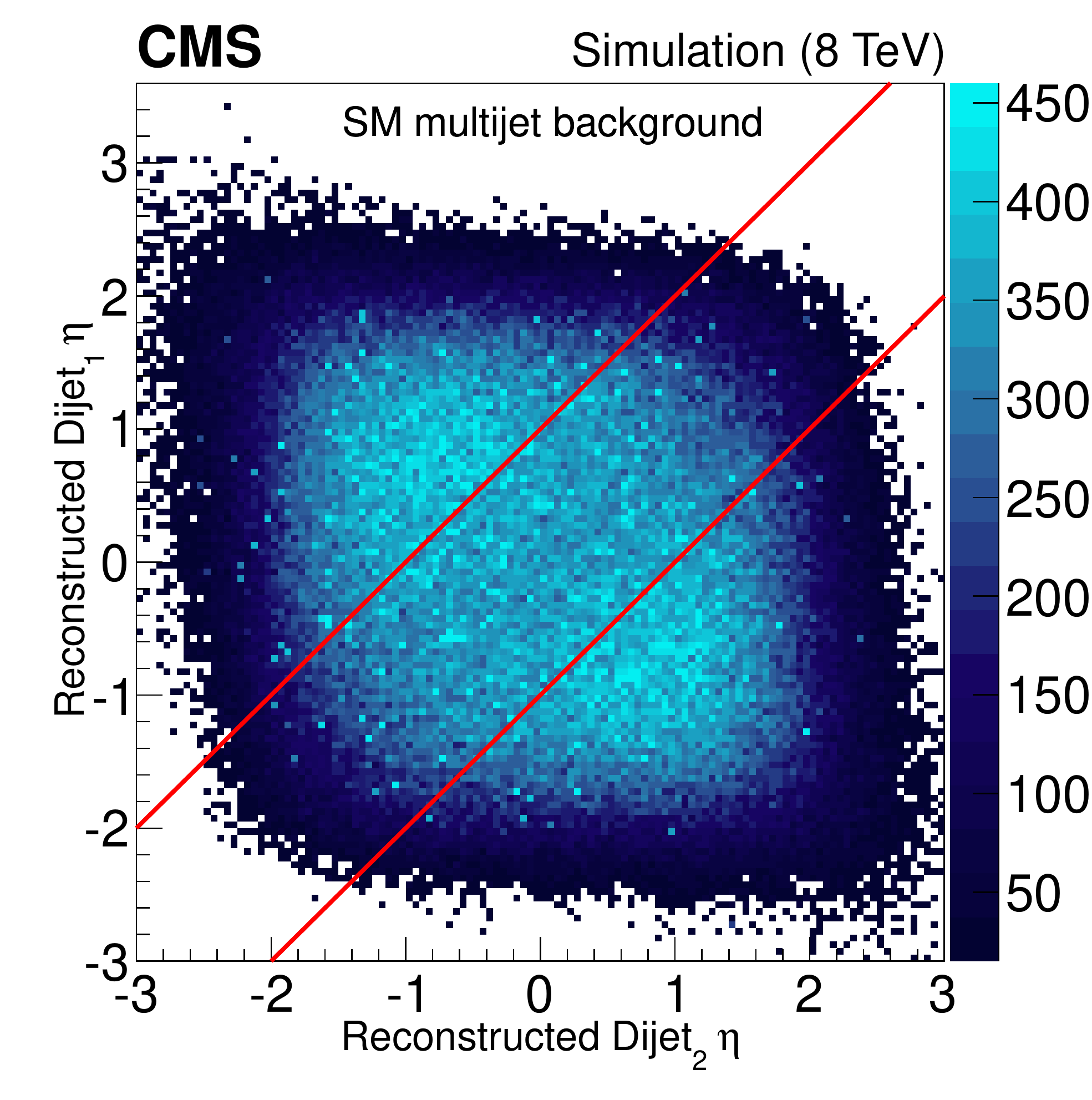}
        \caption{The $\eta$ value for the higher-\pt reconstructed dijet system versus that
                of the lower-\pt dijet system in the selected pair. This distribution is
                shown for 400\GeV top squark (\cmsLeft) and simulated SM multijet
                samples (\cmsRight), with the right hand scale indicating the expected number
                of events per bin. The diagonal lines indicate the optimized region of
                allowed $\Delta\eta_{\text{dijet}}$ values, and events with values falling between
                the two lines pass this requirement.}
\label{fig:d_eta}
\end{figure}

The optimization for the signal selection is performed
as a function of the three kinematic variables described above:
$\Delta m/m_{\text{av}}$, $\Delta\eta_{\text{dijet}}$, $\Delta$,
as well as the fourth jet \pt.
Because the number of expected background events is large,
we use $S/\sqrt{B}$ as the metric for signal optimization, where $S$ and
$B$ are the number of signal and background events, respectively,
and $B$ is determined by using the $m_{\text{av}}$ of simulated SM events.
The values of $S$ and $B$ are set to the number of events within a window
of width $\pm$10\% centered at the generated top squark mass, where
the value of 10\% is roughly twice the expected resolution for signal masses.
We study this metric by evaluating $S$ and $B$ based on events passing
a number of thresholds of each kinematic variable and obtain several
four-dimensional tables, in which a value of $S/\sqrt{B}$ is found for every combination of
the four variables. These tables are produced in the low- and high-mass
search regions, and for the inclusive and heavy-flavor analyses separately.
An example of this is given in Fig.~\ref{fig:sigbkgopt}, where the distribution
for a 500\GeV top squark and for a fit to the simulated SM multijet distribution
are shown for one operating point. The signal shape is bimodal owing to a small
fraction of events with incorrect signal pairings, and the Gaussian peak centered
at the generated mass is the part of the distribution used in the optimization.
The threshold values of the four kinematic variables,
corresponding to maximum values of $S/\sqrt{B}$ in these tables, are taken as a
working point. Because of similar results in this optimization,
the inclusive and heavy-flavor searches use common working points,
with the exception of the heavy-flavor analysis requirement of b tagging.
A summary of the requirements is listed in Table~\ref{tab:allcuts} for both the
low- and high-mass searches. An example of the $\Delta\eta_{\text{dijet}}$ variable
is shown in Fig.~\ref{fig:d_eta}. The correlation between the pseudorapidity
values for the two dijet systems is plotted for both 400\GeV top squark and simulated
SM samples, with the region of allowed values of the $\Delta\eta_{\text{dijet}}$
variable indicated.
For the heavy-flavor search,
we repeat the optimization procedure by using selections based on five
different b-tagged jet configurations:
at least one b-tagged jet in the event, at least one b-tagged jet in the four
highest \pt jets, at least two b-tagged jets in the event, at least two b-tagged
jets in the four highest \pt jets, and at least one b-tagged jet in each of the
two chosen dijet systems. We find that the optimal selection is the requirement
that events contain at least two b-tagged jets among the four highest \pt jets.

After all selection requirements are applied, the fraction of signal
events remaining in the heavy-flavor search ranges from 0.4\% to
1.2\% for the low-mass search and from 0.4\% to
1.6\% for the high-mass search. For the inclusive
search, the fraction of signal events remaining ranges from 1.4\%
to 7.4\% for the low-mass search and from 1.4\% to
6.5\% for the high-mass search. In all scenarios, the leading
efficiency loss is due to the required jet \pt thresholds.
In the data, approximately 20\%
of the selected events passing the high-mass search criteria are
in common with the low-mass search.
\begin{table*}[htbp]
\centering
\topcaption{Summary of the low- and high-mass selection criteria for both the
inclusive and heavy-flavor analyses. For the heavy-flavor analysis,
in addition to the requirements below, at least two of the four highest \pt
jets must be b-tagged.}
\begin{tabular}{c  c  c}
                      & Low-mass search & High-mass search \\ \cline{2-3}
 Mass range             & 200--300\GeV  & $>$300\GeV\\
 Integrated luminosity  & 12.4\fbinv  & 19.4\fbinv \\ \hline
$\Delta m/m_{\text{av}}$ & $<$0.15            & $<$0.15    \\
$\Delta\eta_{\text{dijet}}$ & $<$1.0             & $<$1.0     \\
$\Delta$           & $>$70\GeV          & $>$100\GeV \\
Fourth jet \pt     & $>$80\GeV          & $>$120\GeV \\
\end{tabular}
\label{tab:allcuts}
\end{table*}

\section{Background estimation and systematic uncertainties}
\begin{figure*}[hbt]
\centering
\includegraphics[width=\cmsFigWidthTwo]{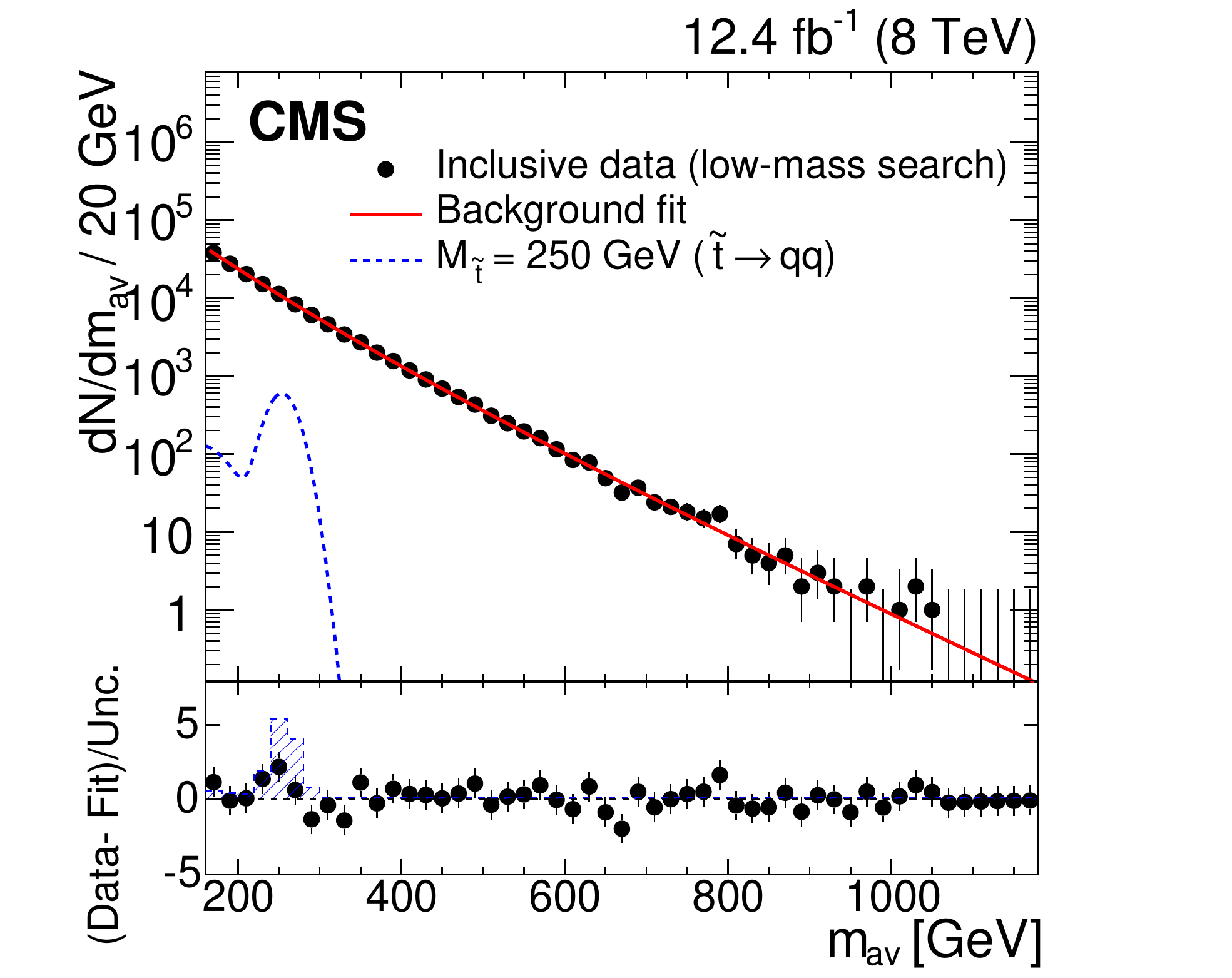}
\includegraphics[width=\cmsFigWidthTwo]{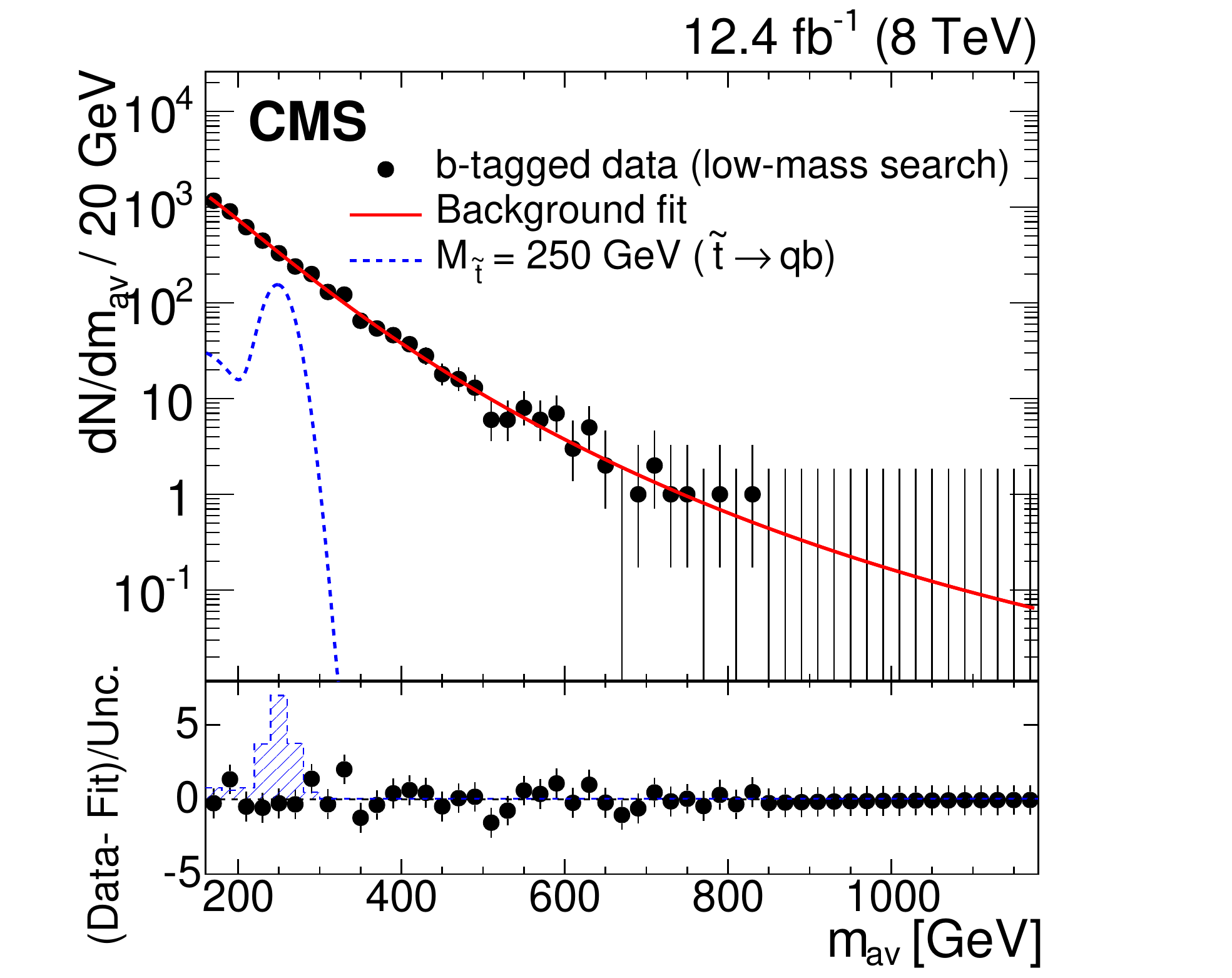}
\includegraphics[width=\cmsFigWidthTwo]{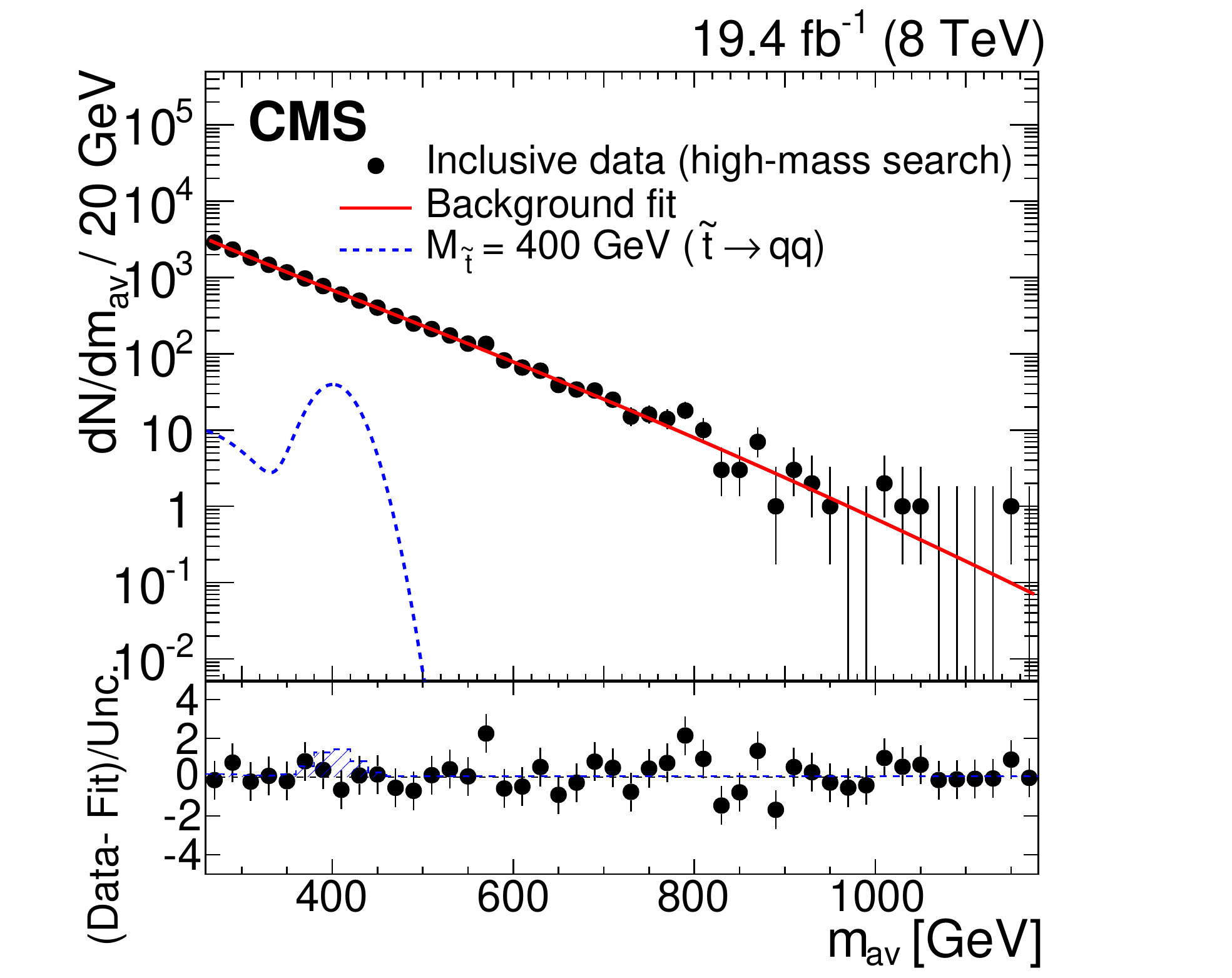}
\includegraphics[width=\cmsFigWidthTwo]{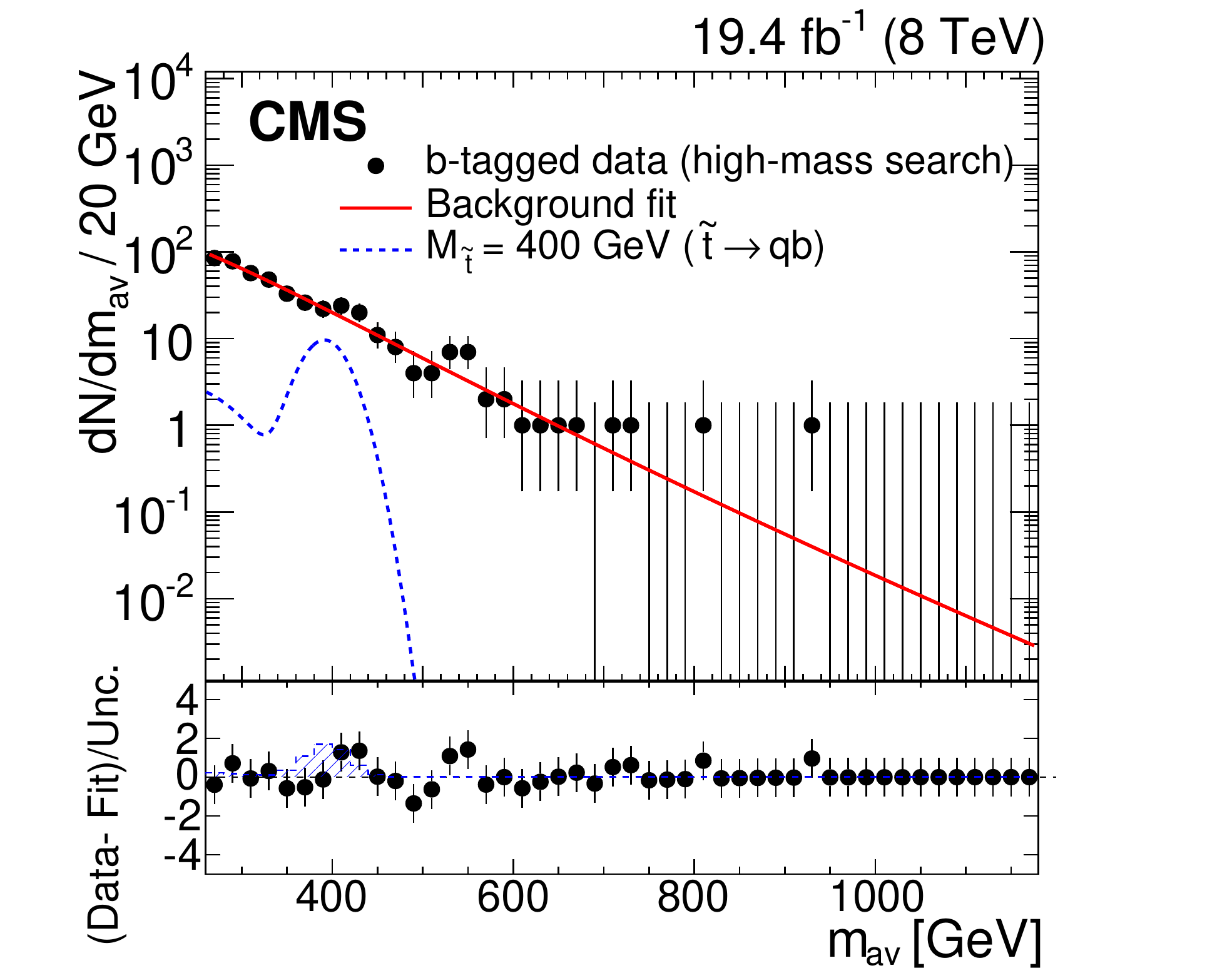}
\caption{The $m_{\text{av}}$ distributions with the superimposed fit from
         Eq.~(\ref{Eq:p4}).
         The events shown satisfy requirements for the inclusive
         searches (left) and the heavy-flavor searches (right) in
         the low-mass (top) and high-mass (bottom) scenarios.
         The expectation
         for the top squark signal is indicated by the blue dashed
         line for the low-mass search ($M_{\PSQt}= 250$\GeV)
         and for the high-mass search ($M_{\PSQt}= 400$\GeV).
         The bottom part of each figure shows the difference in each bin
         between the data and the background estimate divided by the
         statistical uncertainty associated with the data, with the
         shaded region indicating the expected distribution in the
         case of the top squark signal appearing in data. The last
         bin in each $m_{\text{av}}$ distribution also includes
         all overflow $m_{\text{av}}$ events.}
\label{fig:dataplots}
\end{figure*}

The dominant background for this search comes from SM multijet
events. Following a method used previously for similar resonance
searches~\cite{dijetmass8tev,cmsmultijets,gluino2011,gluino2014},
the steeply falling SM background shape is modeled with the use
of a four-parameter function:
\begin{equation}
\frac{\rd{}N}{\rd{}m_{\text{av}}}=p_0 \frac{\left(1 - \frac{m_{\text{av}}}{\sqrt{s}}\right)^{p_1}}{\left(\frac{m_{\text{av}}}{\sqrt{s}}\right)^{p_2 + p_3 \log{\frac{m_{\text{av}}}{\sqrt{s}}}}},
\label{Eq:p4}
\end{equation}
where $N$ is the number of events
and $p_0$ through $p_3$ are parameters of the function. Localized
deviations of the data from the background hypothesis are indications
of a signal, and the fitted data distributions for the four search
scenarios are shown in Fig.~\ref{fig:dataplots}.
The search itself is restricted to the region modeled by the
background parameterization, which begins at 200\GeV for the low-mass
scenario and at 300\GeV for the high-mass scenario.
The agreement of each background fit to its respective mass
distribution is quantified by computing in each bin the difference of the
data and the fit, divided by the statistical
uncertainty associated with the data. These distributions indicate that
no significant deviation is found in any of the four search
scenarios.

The dominant systematic uncertainties that affect the yield originate
from six sources: the imperfect knowledge of the integrated luminosity
(2.6\%)~\cite{lumi}; the simulation of initial-state radiation
(5\%)~\cite{madgraph}; the precision of the jet energy corrections
(1--6.2\%)~\cite{jec}; the jet energy resolution (10\%)~\cite{jec};
the efficiency of b tagging (2\%)~\cite{btagger}; the modeling of the
effect of multiple pp interactions ($<$1.5\%)~\cite{pileupuncert}.
We use log-normal priors to model systematic uncertainties
on the signal, which are treated as nuisance parameters.
To ensure that the choice of background parameterization does not introduce
any bias to the estimate of the background obtained from the fit, studies are
performed to derive the appropriate associated uncertainties. For the choice of function
used to model the background shape, we consider several families of functions as a basis
of comparison: exponentials, power-law functions, and Laurent series. Using a method
previously employed by CMS~\cite{higgsgamma2}, we study the difference in expected
yield in the presence of a signal by using each of these functions instead of the
default one, using simulated SM events as the default background shape as input to
the pseudo-experiments.

For each pseudo-experiment, each of the parameterizations is fit to the fluctuated background shape,
and the largest value of the fractional difference between the alternate fit result and the default
one is calculated for every $m_{\text{av}}$ bin. The mean of the resulting distribution is taken as the
bin-by-bin uncertainty for each alternate parameterization, and the average of the alternate parameterization
uncertainties determines the overall assigned uncertainty. This uncertainty
increases with $m_{\text{av}}$ from 0.3\% to 0.6\% in the
low-mass search range, and from 0.5\% to 30\% in the high-mass search range.

\section{Results}
We set upper limits on the production cross section
using a Bayesian formalism with a uniform prior for
the cross section.
The binned likelihood $L$ can be written as
\begin{equation}
L = \prod_{i} \frac{\mu_{i}^{n_{i}}\re^{-\mu_{i}}}{n_{i}!},
\end{equation}
where $\mu_{i}$ is defined as $\mu_{i} = \alpha N_{i}(S) + N_{i}(B)$ and
$n_{i}$ is the measured number of events in the $i$th bin of $m_{\text{av}}$.
Here, $N_{i}(S)$ is the number
of expected events from the signal in the $i$th $m_{\text{av}}$ bin, $\alpha$ is a
constant to scale the signal amplitude, and $N_{i}(B)$ is the number of
expected events from background in the $i$th $m_{\text{av}}$ bin.
The likelihood is combined with the prior and nuisance parameters,
and then marginalized to give the posterior density for the signal
cross section. Integrating the posterior density to 0.95 of the total
gives the 95\% CL limit for the signal cross section.
The expected limits on the cross section are estimated with
pseudo-experiments generated using background shapes,
obtained by signal-plus-background fits to the data.
Closure tests are performed where a fixed signal is injected,
and these confirm that the presence of signal would not be
hidden in the estimated background.

\begin{figure}[hbt]
\centering
\includegraphics[width=\cmsFigWidthTwo]{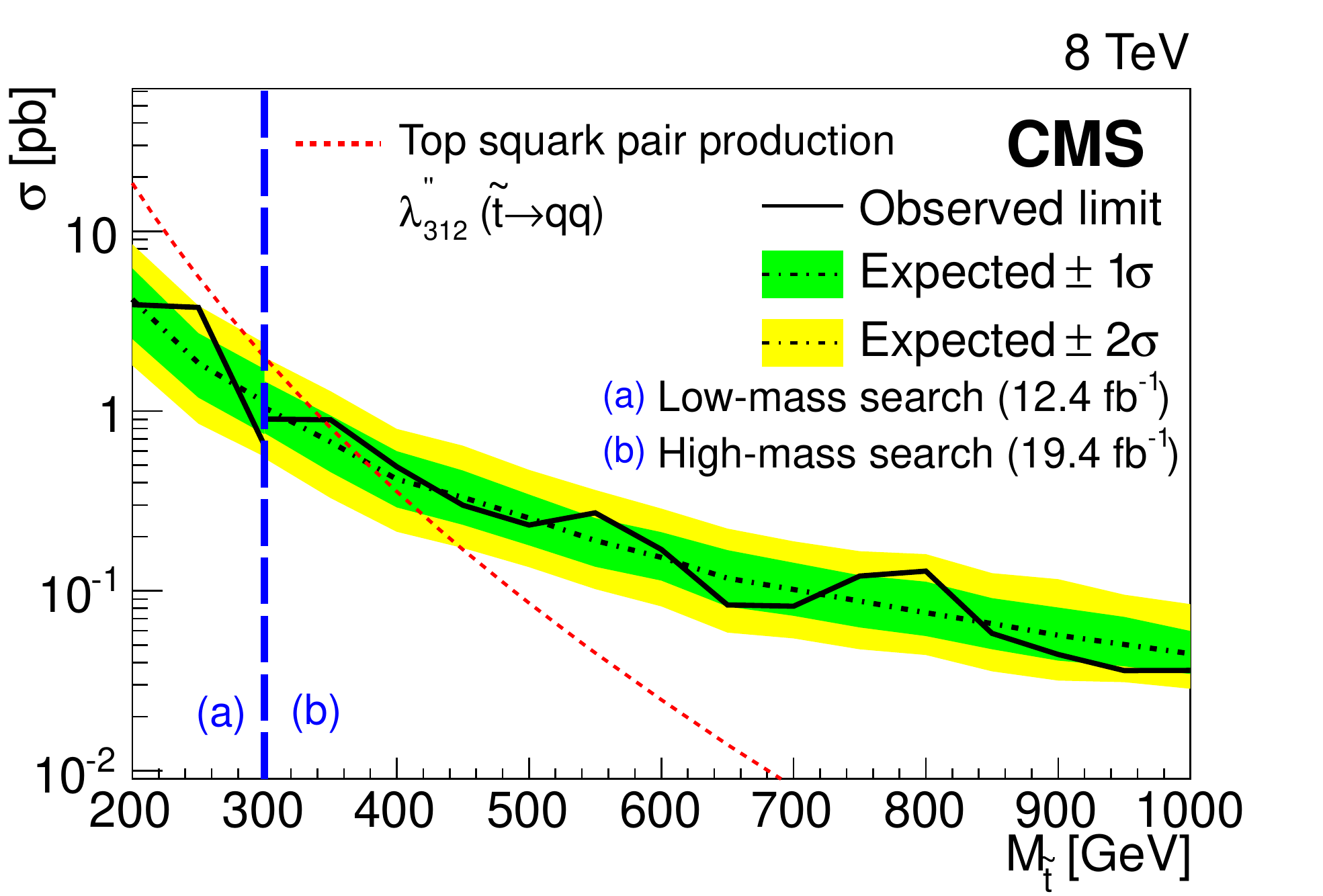}
\includegraphics[width=\cmsFigWidthTwo]{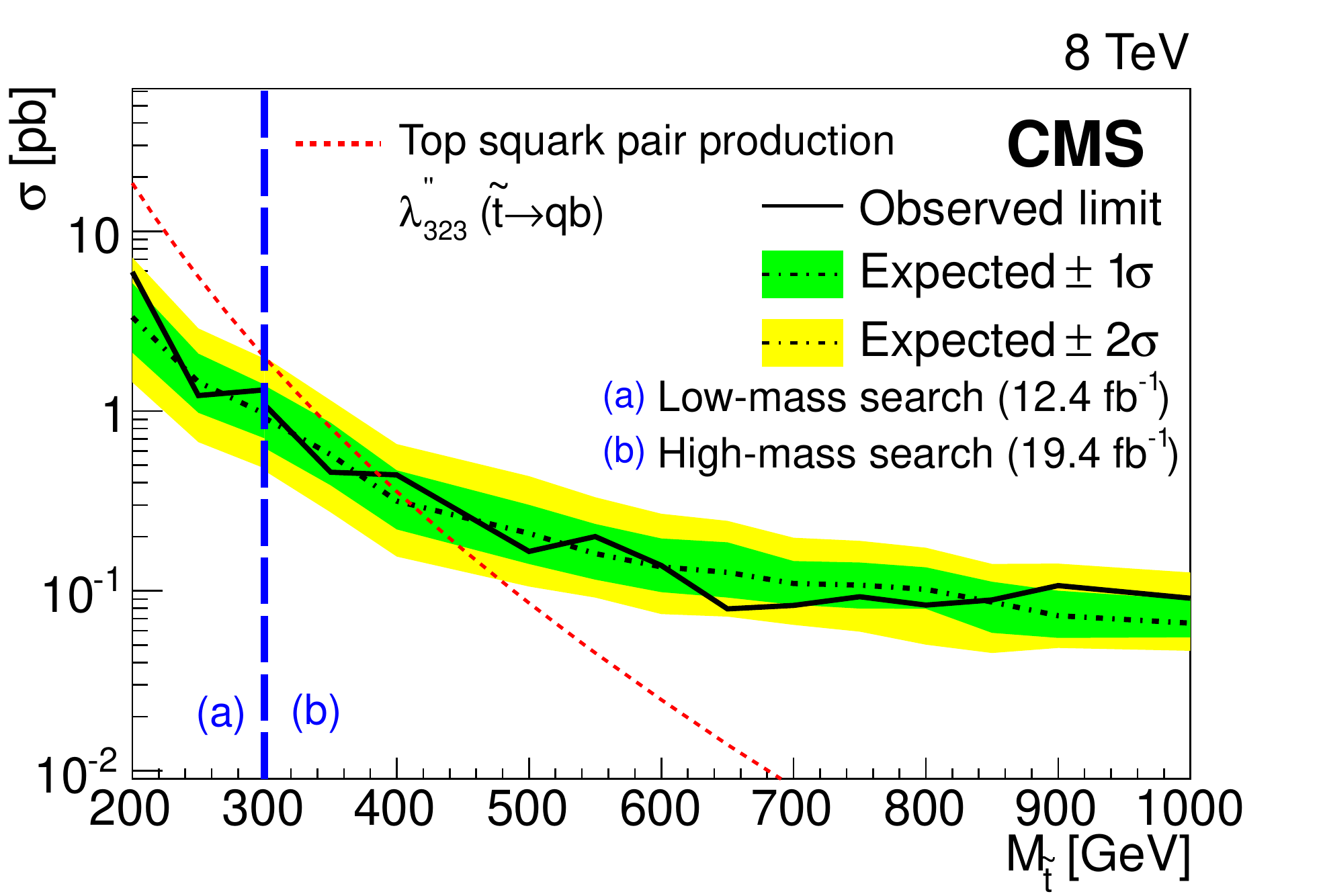}
\caption{Observed and expected 95\% CL cross section limits as a function of top squark mass
         for the inclusive (\cmsLeft) and heavy-flavor (\cmsRight) RPV top squark
         searches based on results from
         the low-mass (a) and high-mass (b) scenarios.
         The dotted red line shows the NLO+NLL
         predictions for top squark production, and the vertical dashed blue line indicates the
         boundary of the limits between the low- and high-mass scenarios.}
\label{fig:Limits}
\end{figure}

\begin{figure}[hbt]
\centering
\includegraphics[width=\cmsFigWidthTwo]{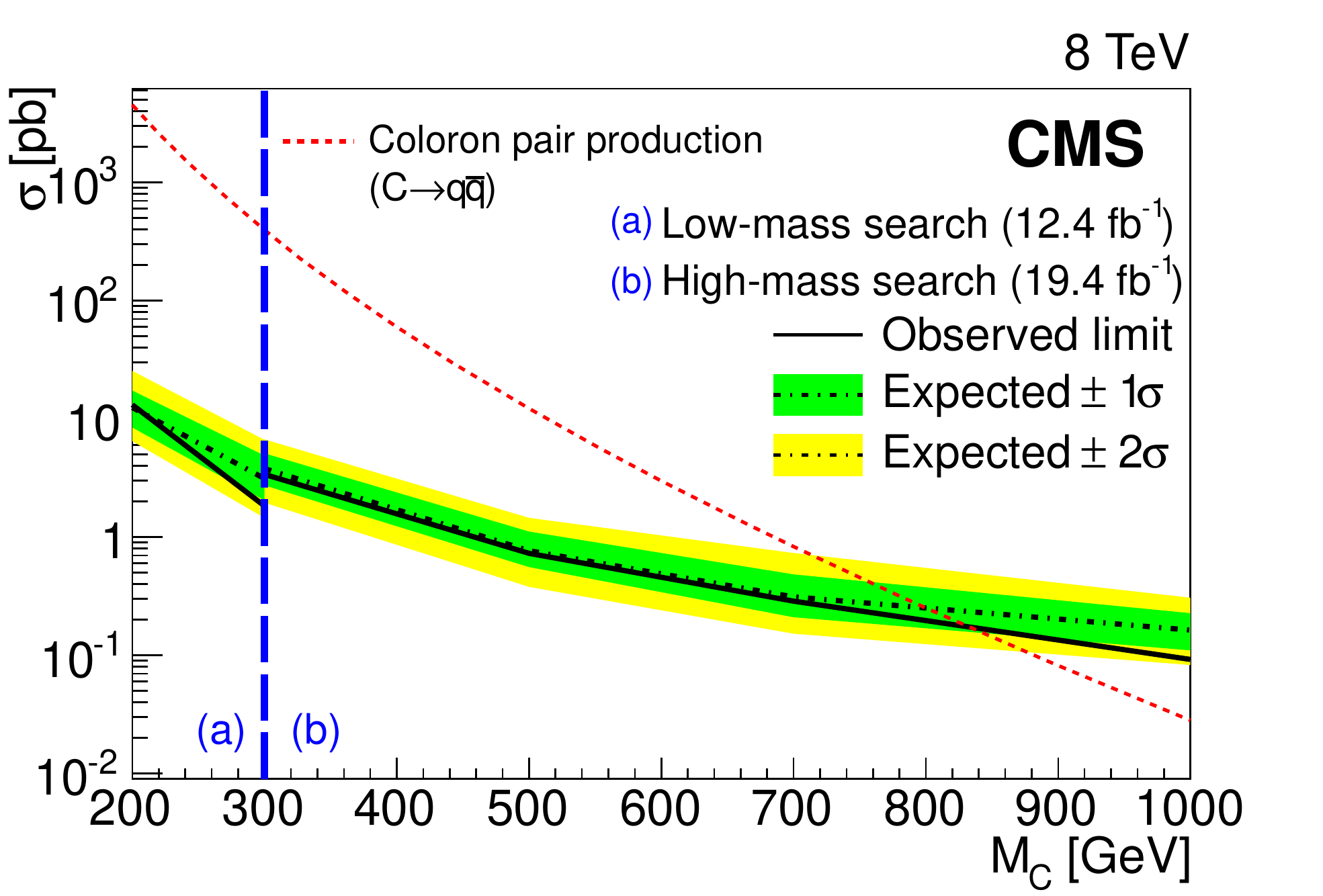}
\caption{Observed and expected 95\% CL cross section limits as a function of coloron mass
         for the pair-produced coloron search based on results from the low-mass~(a) and
         high-mass (b) scenarios. The dotted red line shows the NLO+NLL
         predictions for coloron pair production, and the vertical dashed blue line indicates the
         boundary of the limits between the low- and high-mass scenarios.}
\label{fig:Limitscoloron}
\end{figure}

Figure~\ref{fig:Limits} shows the observed and expected
95\% CL upper limits on $\sigma$, the cross section, and
a dotted red line indicating the NLO+NLL
predictions for top squark production~\cite{susyxsec1, susyxsec2, susyxsec3, susyxsec4, susyxsec5},
where the top squark mass is equal to $m_{\text{av}}$.
The vertical dashed blue line at a top squark mass of 300\GeV
indicates the transition from the low- to the high-mass limits,
and at this mass point the limits are shown for both analyses.
The production of top squarks undergoing RPV decays into light-flavor
jets is excluded at 95\% CL for top squark masses from 200 to 350\GeV.
Top squarks whose decay includes a heavy-flavor jet are excluded
for masses between 200 and 385\GeV.
We exclude the production
of colorons decaying into four jets at 95\% CL for masses between
200 and 835\GeV, as seen in Fig.~\ref{fig:Limitscoloron}.

\section{Summary}
A search has been performed for pair production of heavy resonances decaying
to pairs of jets in four-jet events
from proton-proton collisions at $\sqrt{s}=8$\TeV with the CMS
detector. The distribution in the average mass of selected
dijet pairs has been investigated for localized disagreements between the data
and the background estimate. This method takes advantage of a number of
additional optimized kinematic requirements imposed on the dijet pair.
No significant deviation is found between the selected events and the
expected standard model multijet background.
Limits are placed on the production of colorons decaying
into four jets with a 100\% branching fraction, excluding at 95\% confidence level, masses between
200 and 835\GeV. For this model, these results include first limits
in the mass ranges of
200--250\GeV and 740--835\GeV, extending previous limits~\cite{exo11016}
to lower masses by 50\GeV, and to higher masses by 95\GeV.
Limits are set on top squark pair production through
the $\lambda^{''}_{\mathrm{UDD}}$ coupling to final states with either only
light-flavor jets or both light- and heavy-flavor jets with a 100\% branching fraction.
We exclude at a 95\% confidence level top squark production followed by R-parity violating decays to light-flavor jets
for top squark masses from 200 to 350\GeV and decays to heavy-flavor
jets for masses between 200 and 385\GeV. Both sets of limits are the most stringent such limits to date, and
the first from the LHC for this model of R-parity violating top squark decay.

\begin{acknowledgments}
We congratulate our colleagues in the CERN accelerator departments for the excellent performance of the LHC and thank the technical and administrative staffs at CERN and at other CMS institutes for their contributions to the success of the CMS effort. In addition, we gratefully acknowledge the computing centers and personnel of the Worldwide LHC Computing Grid for delivering so effectively the computing infrastructure essential to our analyses. Finally, we acknowledge the enduring support for the construction and operation of the LHC and the CMS detector provided by the following funding agencies: BMWFW and FWF (Austria); FNRS and FWO (Belgium); CNPq, CAPES, FAPERJ, and FAPESP (Brazil); MES (Bulgaria); CERN; CAS, MoST, and NSFC (China); COLCIENCIAS (Colombia); MSES and CSF (Croatia); RPF (Cyprus); MoER, ERC IUT and ERDF (Estonia); Academy of Finland, MEC, and HIP (Finland); CEA and CNRS/IN2P3 (France); BMBF, DFG, and HGF (Germany); GSRT (Greece); OTKA and NIH (Hungary); DAE and DST (India); IPM (Iran); SFI (Ireland); INFN (Italy); MSIP and NRF (Republic of Korea); LAS (Lithuania); MOE and UM (Malaysia); CINVESTAV, CONACYT, SEP, and UASLP-FAI (Mexico); MBIE (New Zealand); PAEC (Pakistan); MSHE and NSC (Poland); FCT (Portugal); JINR (Dubna); MON, RosAtom, RAS and RFBR (Russia); MESTD (Serbia); SEIDI and CPAN (Spain); Swiss Funding Agencies (Switzerland); MST (Taipei); ThEPCenter, IPST, STAR and NSTDA (Thailand); TUBITAK and TAEK (Turkey); NASU and SFFR (Ukraine); STFC (United Kingdom); DOE and NSF (USA).

Individuals have received support from the Marie-Curie program and the European Research Council and EPLANET (European Union); the Leventis Foundation; the A. P. Sloan Foundation; the Alexander von Humboldt Foundation; the Belgian Federal Science Policy Office; the Fonds pour la Formation \`a la Recherche dans l'Industrie et dans l'Agriculture (FRIA-Belgium); the Agentschap voor Innovatie door Wetenschap en Technologie (IWT-Belgium); the Ministry of Education, Youth and Sports (MEYS) of the Czech Republic; the Council of Science and Industrial Research, India; the HOMING PLUS program of Foundation for Polish Science, cofinanced from European Union, Regional Development Fund; the Compagnia di San Paolo (Torino); the Consorzio per la Fisica (Trieste); MIUR project 20108T4XTM (Italy); the Thalis and Aristeia programs cofinanced by EU-ESF and the Greek NSRF; and the National Priorities Research Program by Qatar National Research Fund.
\end{acknowledgments}
\bibliography{auto_generated}
\cleardoublepage \appendix\section{The CMS Collaboration \label{app:collab}}\begin{sloppypar}\hyphenpenalty=5000\widowpenalty=500\clubpenalty=5000\textbf{Yerevan Physics Institute,  Yerevan,  Armenia}\\*[0pt]
V.~Khachatryan, A.M.~Sirunyan, A.~Tumasyan
\vskip\cmsinstskip
\textbf{Institut f\"{u}r Hochenergiephysik der OeAW,  Wien,  Austria}\\*[0pt]
W.~Adam, T.~Bergauer, M.~Dragicevic, J.~Er\"{o}, M.~Friedl, R.~Fr\"{u}hwirth\cmsAuthorMark{1}, V.M.~Ghete, C.~Hartl, N.~H\"{o}rmann, J.~Hrubec, M.~Jeitler\cmsAuthorMark{1}, W.~Kiesenhofer, V.~Kn\"{u}nz, M.~Krammer\cmsAuthorMark{1}, I.~Kr\"{a}tschmer, D.~Liko, I.~Mikulec, D.~Rabady\cmsAuthorMark{2}, B.~Rahbaran, H.~Rohringer, R.~Sch\"{o}fbeck, J.~Strauss, W.~Treberer-Treberspurg, W.~Waltenberger, C.-E.~Wulz\cmsAuthorMark{1}
% ...
\vskip\cmsinstskip
% ...
1:~~Also at Vienna University of Technology, Vienna, Austria\\
2:~~Also at CERN, European Organization for Nuclear Research, Geneva, Switzerland\\
% ...

\end{sloppypar}
\end{document}